# MULTISCALE GEOMETRIC ANALYSIS OF DYNAMIC WETTABILITY ON COMPLEX, FRACTAL-LIKE, ANISOTROPIC SURFACES

Katarzyna Peta[1,*], Krzysztof J. Kubiak[2], Christopher A. Brown[3]

[1] Institute of Mechanical Technology, Poznan University of Technology, 60-965 Poznan, Poland

[2] School of Mechanical Engineering, University of Leeds, United Kingdom

[3] Mechanical and Materials Engineering Department, Worcester Polytechnic Institute, Worcester, Massachusetts 01609, USA

[*] Corresponding author: katarzyna.peta@put.poznan.pl

Abstract

This study introduces novel insights into the development of procedures for identifying the most relevant scales for observing the interactions of dynamic wettability and surface complexities. The experimental procedures presented for measuring dynamic contact angle hysteresis in multiscale correlation with the geometric characteristics of anisotropic surfaces contribute to a new perspective on measurement practice. In this study, microtexturing with a pyramidal structured abrasive belt is applied for precisely forming area- and length-scale anisotropic surface complexities, and consequently, topographically dependent functional features. The significant role of anisotropic topographies in modeling dynamic wettability behavior is highlighted through multiscale measurement-based analysis. These studies verify the relationship between dynamic wettability and the finest surface microgeometry (microroughness) and also the coarsest texture components (waviness). The size of topographic features, ranging from microroughness to waviness, significantly influences droplet pinning and liquid entrapment. Furthermore, the influence of material hydrophilicity and hydrophobicity on the calculated multiscale relationships is assessed. The results indicated specific scales that best correlate with dynamic wettability, with length- and area-scale complexities of 6.9 μm and 28 μm², respectively. A novel measurement-based approach to scale-dependent surface–functionality interactions offers new insights for designing dynamic wettability on anisotropic surfaces.

## 1. Introduction

Multiscale geometric analyses represent surface topographies as self-similar triangular fractal shapes. The sizes of the triangles define the scale. Smaller triangles cover finer surface geometries. In contrast, larger triangles capture only coarse surface geometries and are associated with larger scales [1]. Surface characterization in the context of multiscale geometry includes the following parameters: relative area (ratio of the calculated areas from triangular tilings at each scale to the nominal surface area) and relative length (ratio of the calculated lengths from approximating the profile with line segments of different lengths to the nominal length segment), as well as area- or length topographic complexity (the slope of the logarithmic relationship of relative area or relative length versus scale, multiplied by -1000) [2]. Multiscale geometric analyses are applicable to surfaces that can be covered with fractal-like shapes. This applies to complex surfaces, most frequently occurring in reality [3]. Surface topographies described by different scales of fractal shapes show different correlation strengths with surface phenomena, including wettability and lubrication [4]. Therefore, one of the important purposes of multiscale geometric analysis is to identify the best scales for observing wetting on complex surfaces [5].

The basis for multiscale characterization of surface wettability is that a liquid droplet interacts discretely with a solid surface along both the contact line and the contact area. Brown analyzed the tribological interactions with the textured base of the ski and noticed local changes in wettability associated with tile inclinations at some significant scales [6]. Triangular tiles were used in this case to visualize and calculate the actual surface area, according to the assumptions of scale-sensitive fractal analysis. It is fundamental that the contact area is characterized by irregularities. Therefore, local contact angles depend on a sufficiently significant inclination of the tiles at certain scales. Despite the possibility of tile inclinations below a certain scale threshold, their interactions with the liquid droplet show no relevant relationship, and these surfaces are considered smooth. The surface irregularities covered by the tiles should be large enough to avoid smoothing the droplet surface caused by surface tension acting at a certain distance from the contact area. The author further stated that interactions related to surface wettability result from the occurrence of numerous discrete, fine-scale interactions.

Wettability studies constitute a scientific challenge due to the difference in scales. Macroscale phenomenon of the behavior of liquid droplets sensitive to surface textures at the nanoscale [7]. Precise surface metrology is important for observing wetting phenomena [8]. Chang et al. concluded that even nanodroplets on nanotextured surfaces have an impact on contact angle hysteresis [9]. Moreover, the indication of the best scales for observing wettability and lubrication phenomena on rough surfaces is described in publications [4,10]. This indicates the importance of the concept of scale in describing topographically dependent tribological functionalities [11].

The functional characteristics of a material largely depend on the topography of its surface [12]. The surface wettability, as one of the basic topographically dependent functionalities, is determined by the contact angle [13]. While static wettability refers to a contact angle of the sessile droplet on a surface, dynamic wettability considers the movement of the droplet and the shift of the contact line [14]. Although the contact angle is considered the basic measure of wettability, it can be supplemented by other characteristics to more broadly describe the dynamic aspects of wettability [15]. The differences between the advancing and receding contact angles for a lubricant droplet in motion determine the dynamic contact angle hysteresis [16]. This phenomenon is an indicator of the interfacial properties related to friction and adhesion of liquid droplets on the surface [17].

The dynamic contact angle hysteresis, closely related to dynamic wettability, is important in, for example, self-cleaning surfaces [18], aerodynamic and hydrodynamic surfaces [19,20], corrosion-resistant surfaces [21], implant lubrication with biological fluids [22], or heat transfer applications [23]. Jaikumar et al. [24] proved that modifying the contact angle hysteresis improves the heat transfer efficiency. This assumption can be effectively determined by experiments or computational fluid dynamics, as demonstrated by Ding et al [25]. Foulkes et al. proposed a nanoscale surface leading to higher heat dissipation coefficients in hot areas while maintaining the required heat transfer efficiency.

Particularly, dynamic contact angle hysteresis strongly correlates with hydrodynamic lubrication [26,27]. The behavior of the lubricant on the surfaces is important in terms of minimizing friction and wear of the surfaces in contact [28]. The sliding angle is important to determine fluid drainage, although in tribology systems, it is used to entrap lubricant [29]. Lubrication depends on the texture of the surface, determining the ability to trap the lubricant in the valleys of the surface or in the spaces between the asperities in the case of irregular surfaces. The entrapment of the lubricant allows for its regular self-supply to the surfaces in contact [30] and modeling the performance of tribological systems [31]. Wettability is also affected by the surface’s lubricant absorption, which is particularly relevant for materials such as natural fibers and high-absorption polymers [32]. Environmental variability (e.g., temperature, humidity, air flow, and other external factors) is particularly important when measuring the functional properties of surface topography [33,34].

The behavior of liquid drops on chemically and topographically different surfaces can be modeled by the principles of dynamics at the three-phase contact line [35]. The contact angle hysteresis can be investigated numerically by the lattice Boltzmann method [36]. Wang et al. performed the first studies on contact angle hysteresis using this method. The authors modeled the dynamic movement of liquid drops for three classic wetting cases, for a perfectly smooth, rough, and chemically inhomogeneous surface [37]. Other work has also considered microporous surfaces and the drainage process [38], as well as anisotropic lubrication of both hydrophilic and hydrophobic surfaces [39]. The flow of liquids over surfaces becomes increasingly complex in wavy or irregular geometries. Surface features affect wettability, the formation of boundary layers, and overall flow stability [40].

Wettability is determined by the chemical composition of the material and surface texture [41]. The literature presents a comparison of static contact angles of various types of materials, including: aluminum, titanium, iron, copper alloys, as well as ceramics and polymers. Regardless of the texture, ceramics are characterized by the highest wettability in this group, and aluminum alloys by the lowest [42]. These differences result from different chemical compositions of the materials and their surface free energy. Surface texture, in turn, influences wettability depending on its different components: roughness, waviness, and lay, which are sequentially presented from the shortest to the longest spatial wavelengths. Roughness and waviness influence wettability [43], while lay is related to the least impact [44]. To extract texture components, S- or L-filtering is used (ISO 25178) [45]. The S-filter attenuates shorter wavelengths, while the L-filter attenuates longer wavelengths. Additionally, by setting certain cutoff values, it is possible to extract only roughness or waviness [46]. Moreover, Ahmad et al. addressed shape-dependent analytical behavior in geometric mappings by defining appropriate boundary conditions [47]. The decomposition of surface texture components facilitates the investigation of geometric features and corresponding wettability across different scales.

The fabrication of textures that modify wettability, including hydrophilic and hydrophobic surfaces, is a current focus in surface engineering. Nanotexturing of the surface resulted in a small contact angle hysteresis on the condenser surface [48]. Chang et al. studied the effect of patterned surfaces on contact angle hysteresis in a microchannel heat transfer system. The authors observed differences in the contact angle hysteresis when linear surface defects or wear occurred [49]. Some engineering applications require anisotropic wetting and lubrication properties, especially important in heat transfer systems and fluid transport across a surface [50]. Anisotropic wetting and lubrication properties can be created by nano- or micro-texturing the surface in the form of grooves, changing contact angle hysteresis in directions parallel and perpendicular to the grooves [51]. The occurrence of multiple metastable states between the Wenzel and Cassie-Baxter models can be related to grooved surfaces. The advancing and receding contact angles for grooved surfaces depend on the viewing angle [52]. Li described that increasing the spacing between the grooves and reducing their width results in a decrease in the contact angle hysteresis [53].

Abrasive belt grinding is an efficient and precise technique for finishing material surfaces [54]. The abrasive belt's flexibility ensures effective adaptation to the surface geometry, and its extended length facilitates improved heat dissipation, leading to a lower temperature in the grinding zone [55]. A specific type of abrasive belt is one with a pyramidal structure, which offers improved longevity due to the vertical, multi-layer arrangement of abrasive grains in the matrix. Here, self-sharpening material exposes new, sharp abrasive grains as the belt wears, improving the surface finish of the material. Material removal efficiency is closely related to the degree of wear of the pyramidal structures of the abrasive belt [56]. Through controlled and precise material removal, abrasive belts create surface textures that can affect the functional features of the surface.

In recent years, there has been increased attention on multiscale measurement-based techniques for interpreting functional phenomena that depend on surface topography. Despite significant progress in this field, many aspects have not yet been thoroughly investigated. The literature mainly describes multiscale correlations of topographic complexities with static contact angle [4–6,10], dynamic contact angle of isotropic surfaces [57], or microwear [11,58,59]. The studies presented in the literature converge on the conclusion that topographic complexity is an important determinant of liquid behavior on surfaces. He et al. demonstrated that the topographic complexity of a surface increases with the resolution of the observation scale [60]. This relationship can pose a challenge for modeling surface wettability. Chen et al. specifically addressed these challenges in the context of the accuracy of predictive models based on the traditional Wenzel and Cassie–Baxter wettability theories. The authors highlighted the importance of accounting for surface complexity across multiple scales when modeling liquid behavior on surfaces [61]. Armstrong et al. emphasizes that different scales of surface geometric features can affect wettability on a macro scale [62]. Tan et al. highlighted the possibility of extracting surface topography at multiple scales, with each scale contributing a distinct influence on wettability [63]. Peta et al. demonstrated that the strongest correlations between static contact angle and surface complexity occur at the finest scales [4]. In subsequent studies on dynamic wettability of isotropic surfaces, it was demonstrated that the dynamic contact angle hysteresis correlates most strongly with surface complexity at scales ranging from 309 $\mu m^2$ to 756 $\mu m^2$ [57].

Thus, a significant research gap can be identified, which includes the following aspects:

- Abrasive belt grinding has not been considered in terms of the generated multiscale geometries affecting dynamic wettability.
- Dynamic contact angle hysteresis of anisotropic surfaces, which addresses the dynamic aspects of wettability and lubrication, has not yet been considered relative to the scale of observation.
- Dynamic behavior of liquids, which is more natural for functional surfaces, is rarely explored; most publications focus on the static contact angle [36,37,64].
- Grooved, anisotropic textures are characterized by a variety of geometries (depth, width, shape, slope), while conventional parameters for topography characterization provide an incomplete description of this geometry, thus not comprehensive in correlation with wettability and lubrication.

These studies introduce new insights into dynamic surface wettability in a multiscale aspect. The main contributions of this work are as follows:

- Multiscale measurement-based correlations of dynamic contact angle hysteresis and topographic complexity of anisotropic surfaces.
- Indicating the differences in dynamic wettability between hydrophilic and hydrophobic surfaces and the applicability of multiscale analyses to both types of materials.
- Identifying the best scales for observing dynamic wettability.
- Describing anisotropic, grooved textures with area- and length-scale parameters as capable of correlating with dynamic wettability.

## 2. Materials and methods

The following materials were analyzed in these studies:

- aluminium alloy 7064 with hydrophobic properties (contact angle > $90^o$),
- fluorophlogopite mica ceramics in borosilicate glass with hydrophilic properties (contact angle < $90^o$) and chemical composition: 46 wt% $SiO_2$, 17 wt% MgO, 16 wt% $Al_2O_3$, 10 wt% $K_2O$, 7 wt% $B_2O_3$, 4 wt% F.

Fluorophlogopite mica ceramics in borosilicate glass and aluminium alloy 7064 are selected due to their contrasting surface wettability properties. The ceramics provide hydrophilic behavior, while the aluminum alloy exhibits hydrophobic characteristics. This enables a study of dynamic wettability across materials with different surface chemistry. Therefore, these studies investigate dynamic wettability on hydrophobic (aluminum alloy) and hydrophilic (ceramic) surfaces, and explore whether the best observation scales can be generalized across both material types.

Surfaces were initially polished to allow for later groove texturing. The surface preparation included polishing with sandpapers with grits: 400, 600, and 2500, respectively. Then, the grooves were textured with a Trizact (3M, Maplewood, USA) abrasive belt with 600 grit. Abrasive belts create grooved anisotropic surfaces using evenly oriented pyramids of bond with abrasive material (ceramic aluminum oxide). While the pyramids wear, new and sharp abrasive grains constantly appear. The abrasive belts had the same geometric shape of pyramids, and a gradation of the abrasive material embedded in the bond. In contrast to conventional abrasive materials that tend to gouge and plough, Trizact abrasive belts have self-sharpening features that maintain sharp pyramids for precise surface texturing (Fig. 1). The dimensions of the grooves are closely related to the dimensions of the pyramids on the abrasive grinding belt. The grooves were textured by moving the workpiece along one direction of the abrasive belt.

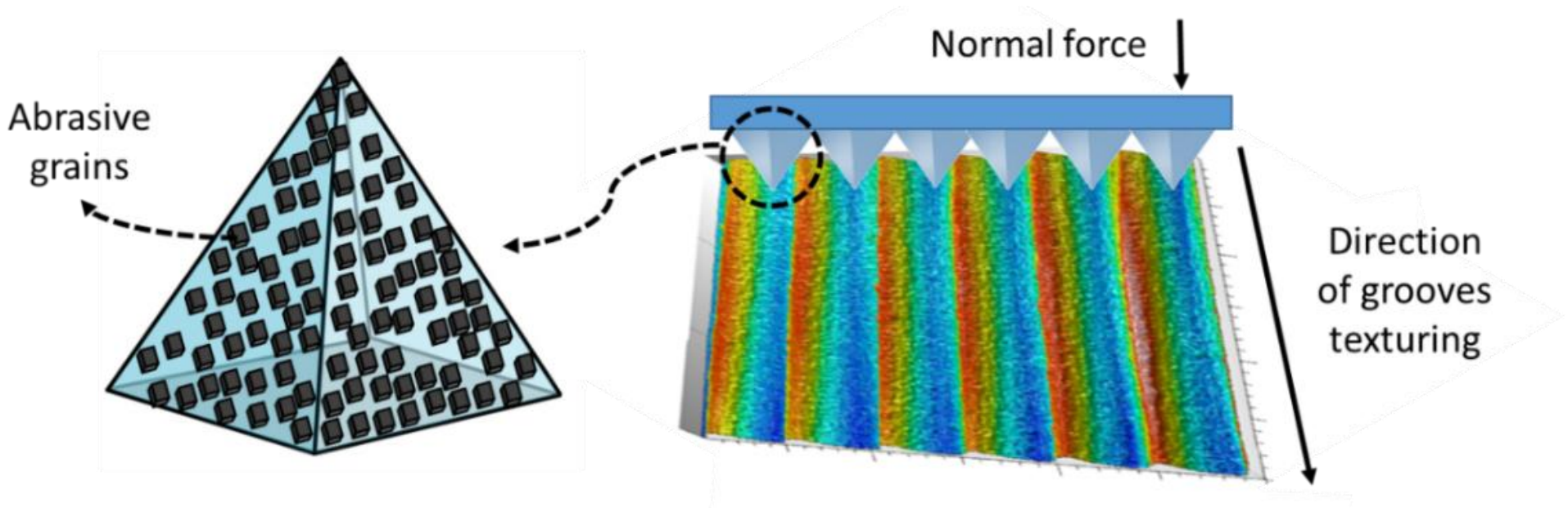

Fig. 1. Graphic visualization of the grinding process with an abrasive belt.

Surface texturing resulted in the formation of ten surfaces for further research (Fig. 2):

- A1 – polished aluminium alloy,
- A2-A5 – aluminum alloy grooved texture,
- C1 – polished ceramics,
- C2-C5 – ceramic grooved texture.

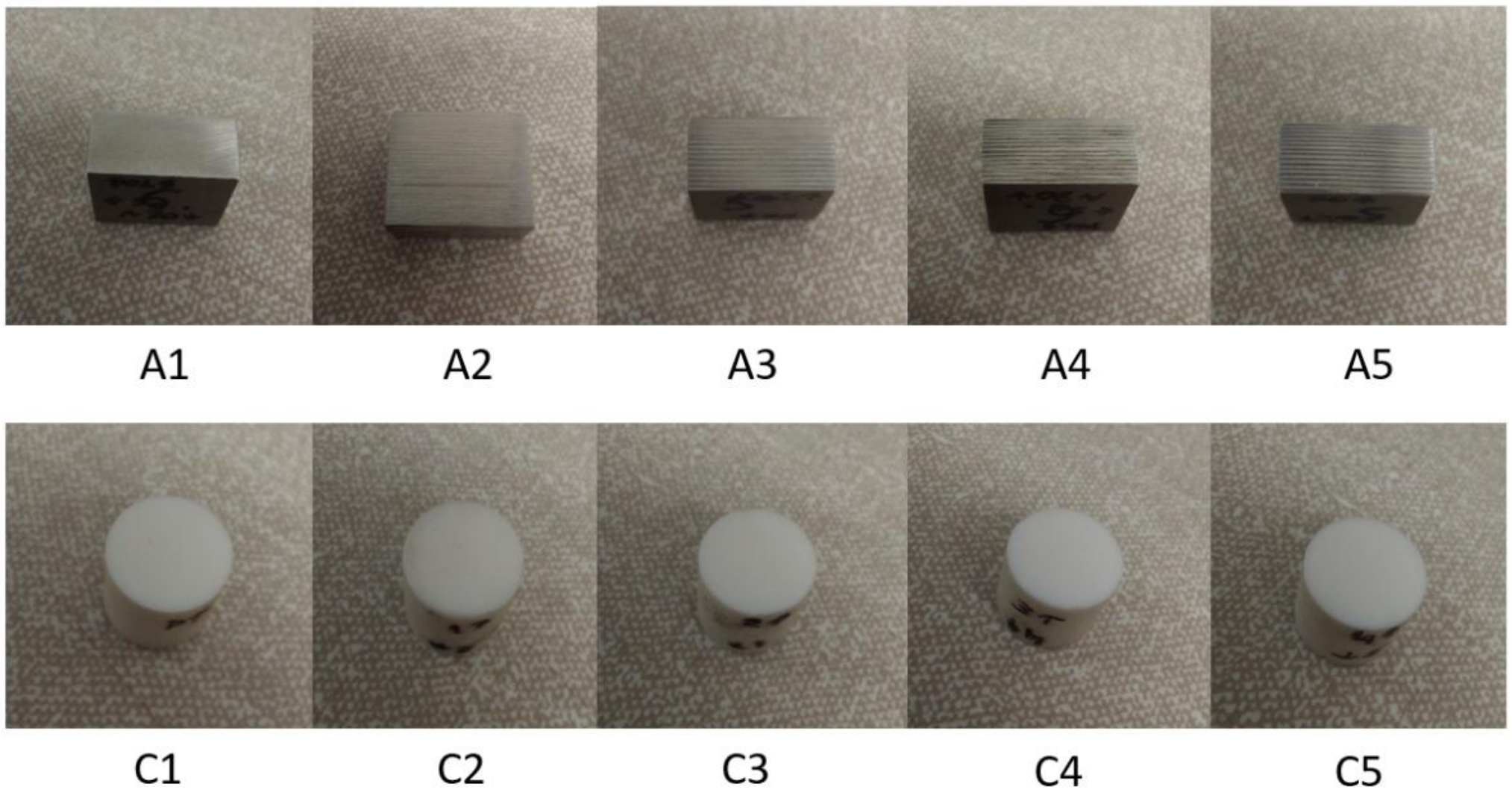

Fig. 2. Textured surfaces of aluminum alloy (A1-A5) and ceramic (C1-C5).

Preparation of the surfaces included cleaning with compressed air (15 s) and a bath in the glass with liquids: deionized water (2 min), isopropyl alcohol (2 min), and acetone (1 min). All surfaces were measured right after surface preparation. Therefore, the time between surface cleaning and both surface topography and wettability measurements was no longer than 5 min.

A Bruker Alicona InfiniteFocus G5 optical 3D microscope (Graz, Austria) was used for topographic characterizations. The measurement settings included 10x magnification, vertical resolution 0.1 µm, and lateral resolution 2.5 µm. The measured topographies were processed in the MountainsMap software from DigitalSurf (Besançon, France), through the following steps: surface leveling, thresholding, removing outliers, and filling in non-measured points. The topographies were characterized by conventional (ISO 25178 [45]), along with multiscale (ASME B46.1 appendix K [65]) parameters. The primary extracted surfaces were filtered to compare the correlation of texture components and the dynamic contact angle hysteresis. For this purpose, Gaussian filtering was used, and a nesting index of 0.25 mm was chosen to attenuate the shortest wavelengths, reducing measurement noise.

Dynamic wettability measurements were performed perpendicular (PR) and parallel (PL) to the grooves on an optical goniometer with the table tilted in the range of 0º-90º. The table was tilted stepwise in 15° increments, from 15° up to 90°, to monitor droplet motion and evaluate dynamic wetting behavior at increasing angles. The needle is attached to a motorized linear stage to facilitate the placement of small droplets on the surface by lowering and retracting the needle. The waiting time of 2 s between successive tilting angles were intended to stabilize the droplet in a given position. The liquid droplet used in the studies was deionized water with a volume of 3 µl. The wettability measurement parameters were selected to ensure repeatable and reproducible characterization of dynamic wettability behavior across materials. Parameters such as droplet volume, deposition method, table tilt angle, and measurement timing were chosen based on experimental experience and literature review to accurately capture the interactions between liquid and solid surface. Here, the results were calculated as an arithmetic mean based on five repetitive wettability measurements on textured surfaces. Wettability measurements were performed at a room temperature of 22°C and a relative humidity of 40%. Dynamic contact angle hysteresis measurements were performed automatically, according to a developed script. The sequence of droplet images was captured using a camera operating at a frame rate of 30 frames per second (30 fps).

*Script of the dynamic contact angle hysteresis measurements*

LT100; *Light On 100%*
DO3; *Drop Out 3µl*
WT2000; *Wait 2000 ms*
MD300; *Move Down the needle 300 steps (3 mm)*
MU300; *Move Up the needle 300 steps (3mm)*
WT2000; *Wait 2000 ms*
TL45; *Table tilt 45º*
WT2000; *Wait 2000 ms*
TL60; *Table tilt 60º*

WT2000;		*Wait 2000 ms*
TL75;	*Table tilt 75º*
WT2000;		*Wait 2000 ms*
TL90;	*Table tilt 90º*
WT2000;		*Wait 2000 ms*
TL0;		*Table tilt back to 0º*

Dynamic wettability included determining the static contact angle, advancing contact angle, receding contact angle, contact angle hysteresis, and sliding angle. An overview of the experimental setup is shown in Fig. 3. Wettability measurements were conducted for a liquid droplet moving perpendicular (PR) and parallel (PL) to the grooves.

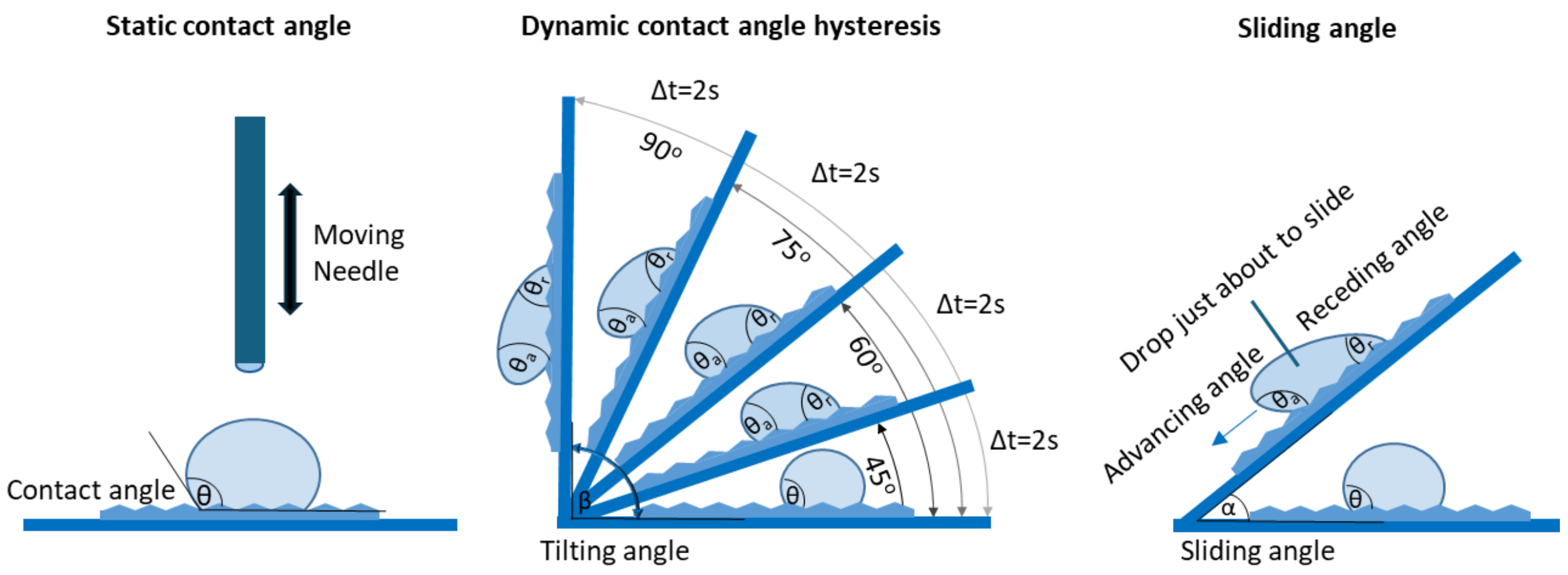

Fig. 3. Schematic view of the experimental setup: dynamic contact angle hysteresis (difference between advancing and receding contact angles), sliding angle, and static contact angle.

Multiscale geometric analyses were applied to correlate surface topographic complexity with dynamic contact angle hysteresis at scales between sampling distance and field of view, i.e., 0.58 $\mu m^2$–1,318,420 $\mu m^2$ for area scales and 1.32 μm–1,624 μm for length scales. Therefore, the textured surfaces were characterized in terms of topographic complexity at the area scale (Asfc) and length scale (Lsfc) in the directions perpendicular and parallel to the grooves, and then correlated with dynamic contact angle hysteresis (CAH). Here, these correlation strengths were calculated based on the Pearson correlation coefficient (r) across the scales considered. Multiscale geometric analyses identified the scales of best positive and negative correlations (r), tending to +1 or -1, between topographic complexities and dynamic contact angle hysteresis. A range of weakest correlation scales close to 0 was also identified. The range of numerical measurement scales can also be presented descriptively, classifying them into three groups: micro-, meso-, and macro-scale. The first group corresponds to the microscale (length: 1.32–100 μm; area: 0.58–1,000 $\mu m^2$), the second to the mesoscale (length: 100–1,000 μm; area: 1,000–1,000,000 $\mu m^2$), and the third to the macroscale (length: 1,000–1,624 μm; area: 1,000,000–1,318,420 $\mu m^2$).

## 3. Results

3D surface images, along with conventional height and hybrid topographic characterization parameters (Sa, Sq, Ssk, Sku, Sp, Sv, Sz, Sdq, Sdr) are presented for the abrasive belt in Fig. 4, textured aluminum alloy in Fig. 5, and textured ceramics in Fig. 6. The texture grooves are closely related to the pyramid shape of the abrasive belt.

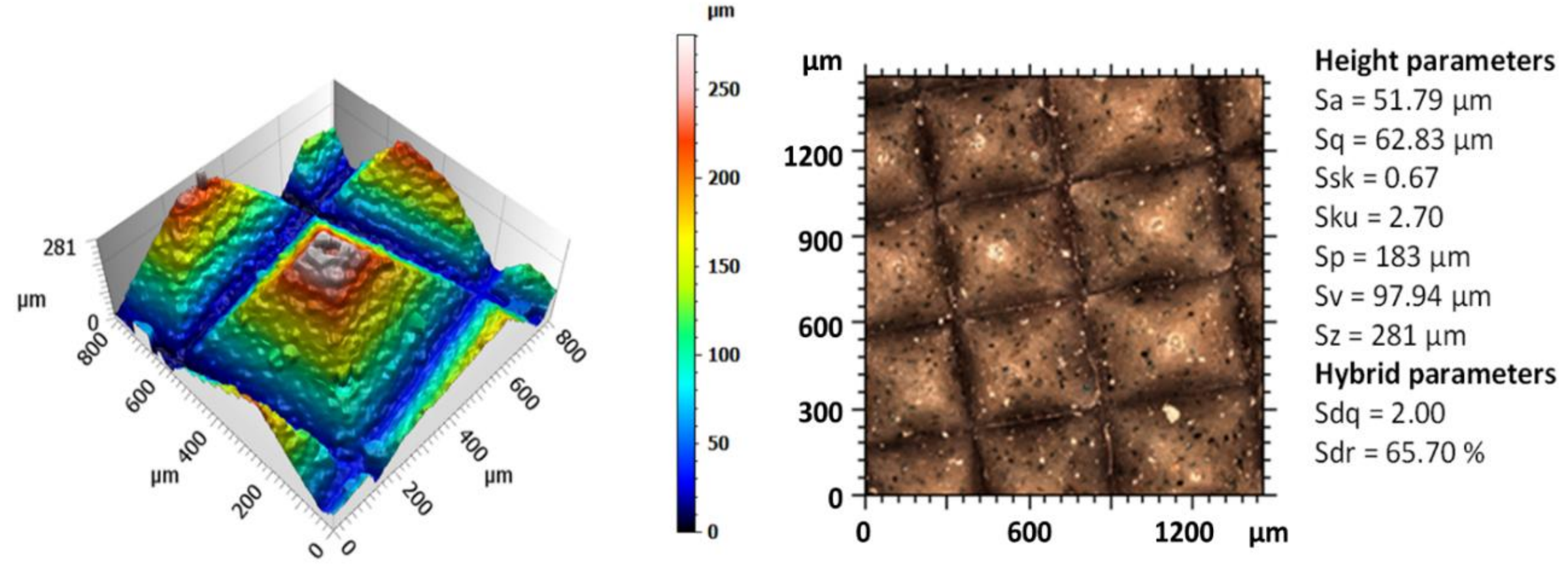

Fig. 4. 3D surface images for abrasive belt.

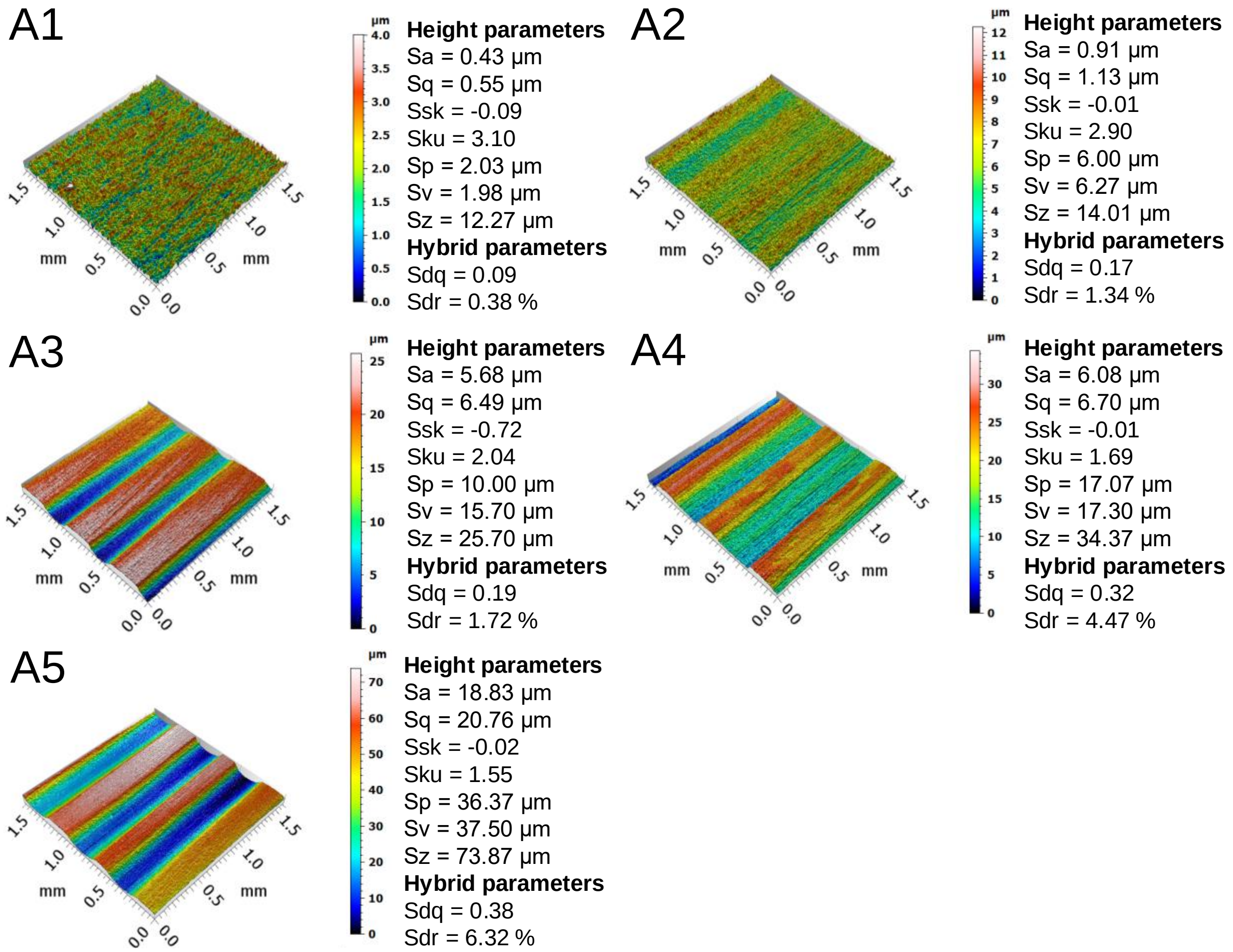

Fig. 5. 3D surface images for polished (A1) and textured surfaces (A2-A5) of aluminum alloy.

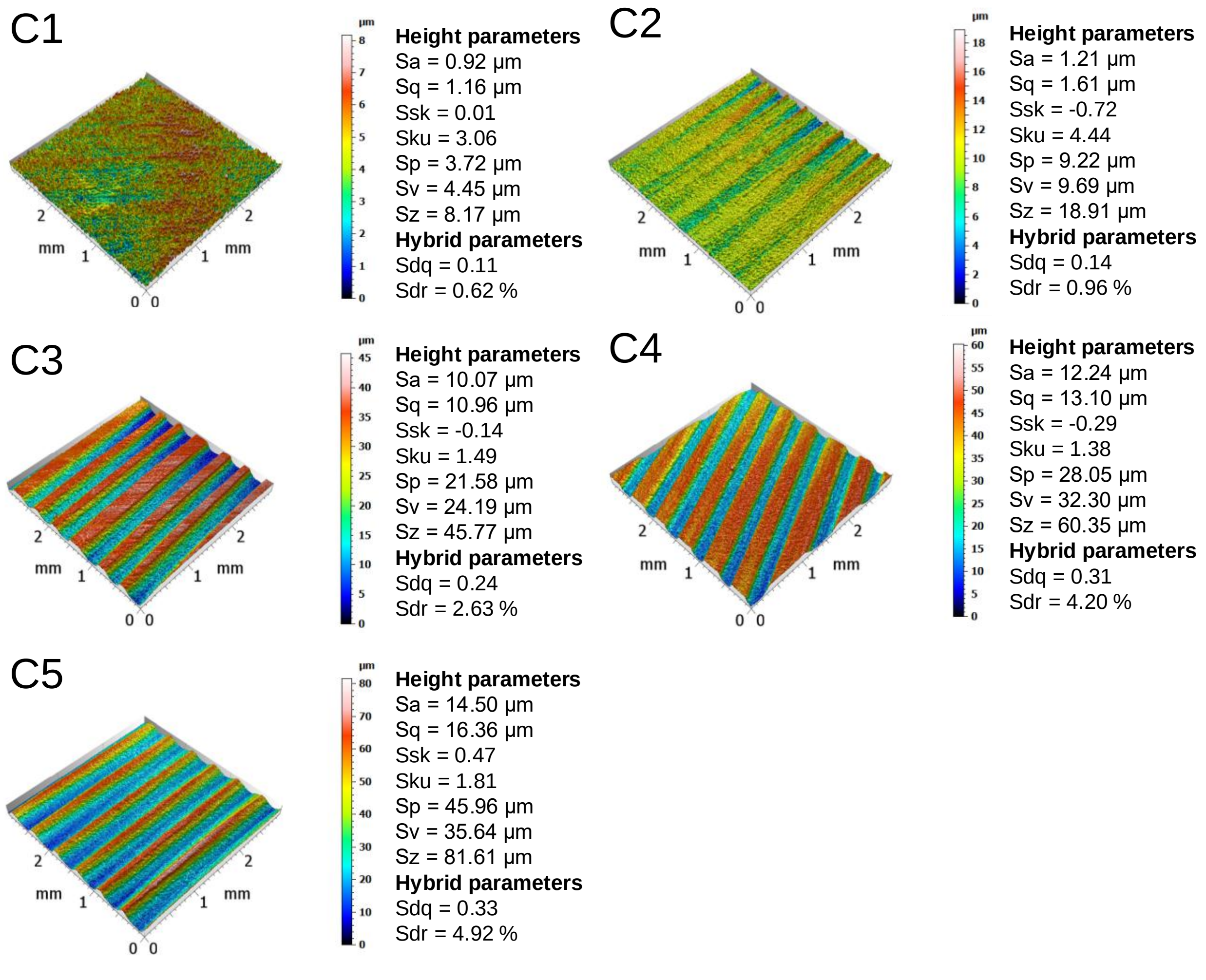

Fig. 6. 3D surface images for polished (C1) and textured surfaces (C2-C5) of ceramic.

Multiscale surface characterizations are presented in Fig. 7. These analyses include: the relative area (RelA) of the textured surfaces (Fig. 7a,b), the area-scale topographic complexity (Asfc) (Fig. 7c,d), the relative length (RelL) of the textured surfaces in both directions parallel (PL) and perpendicular (PR) to the grooves (Fig. 7e,g,i,k), and the length-scale topographic complexity (Lsfc) (Fig. 7f,h,j,l).

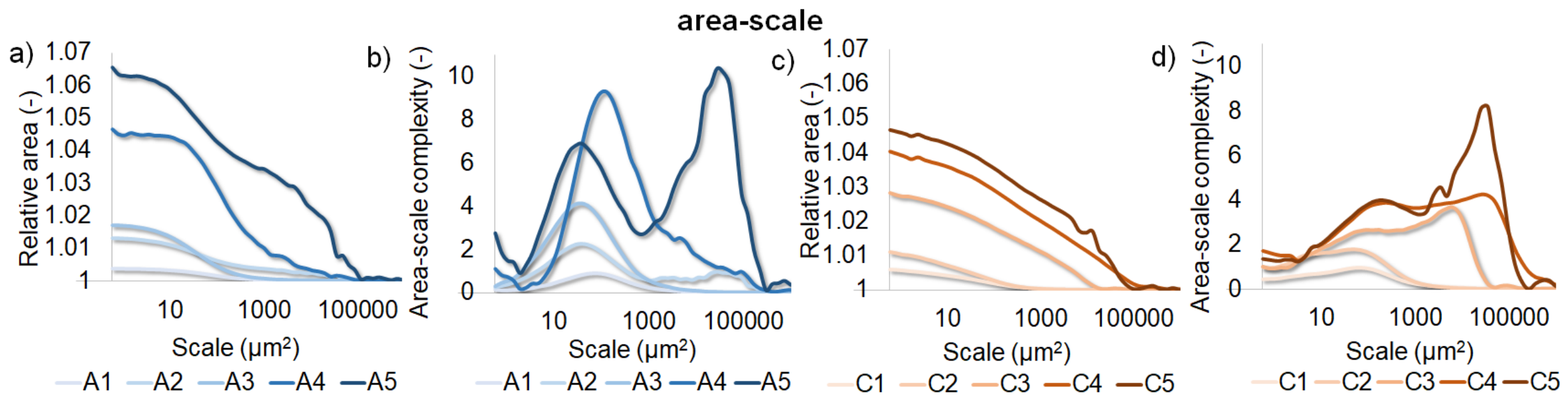

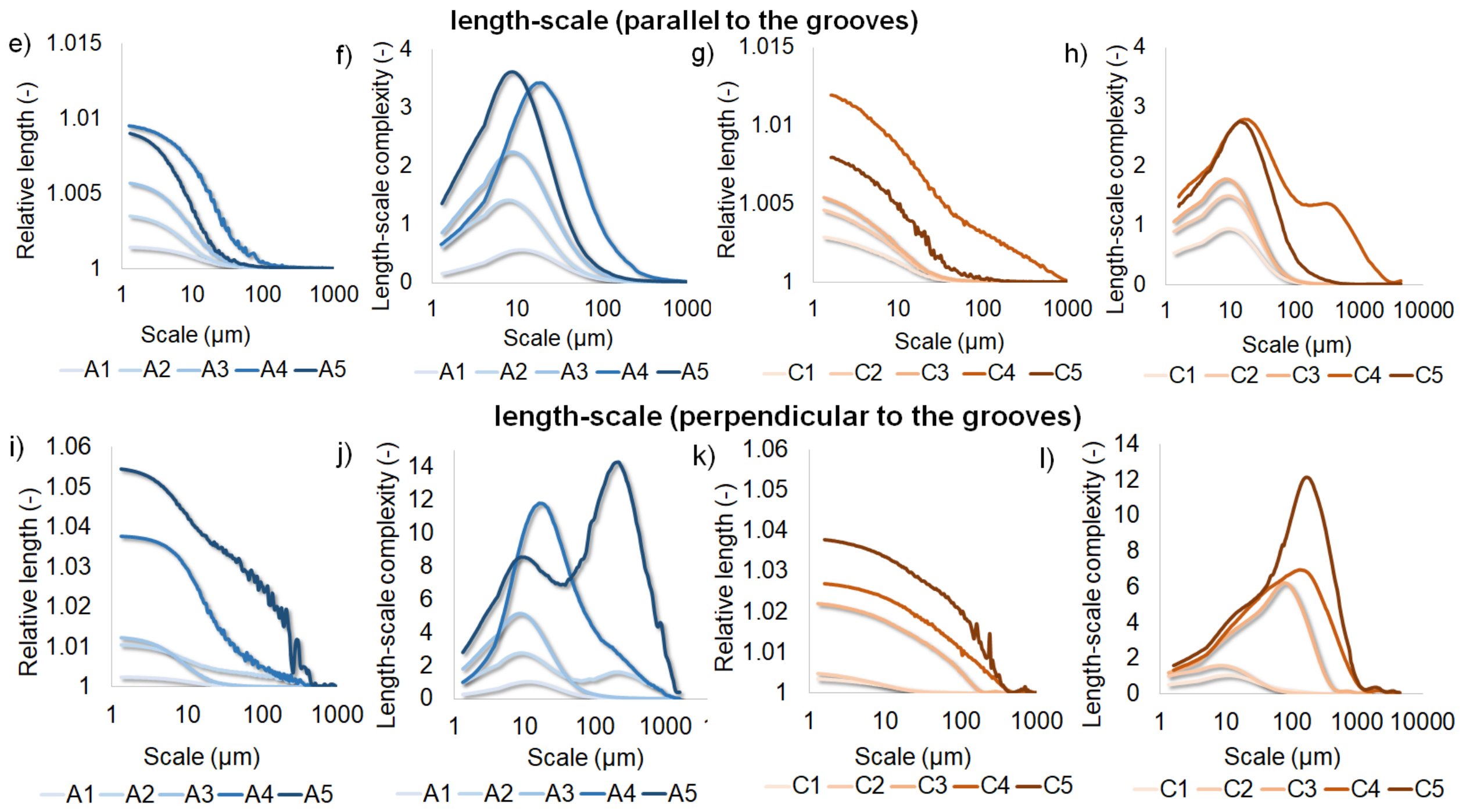

Fig. 7. Multiscale parameters — relative area (RelA), relative length (RelL), area-scale topographic complexity (Asfc), and length-scale topographic complexity (Lsfc) — for aluminum alloy surfaces A1–A5 and ceramic surfaces C1–C5: a) RelA for A1-A5, b) Asfc for A1-A5, c) RelA for C1-C5, d) Asfc for C1-C5, e) RelL for A1-A5 (PL), f) RelL for C1-C5 (PL), g) Lsfc for A1-A5 (PL), h) Lsfc for C1-C5 (PL), i) RelL for A1-A5 (PR), j) RelL for C1-C5 (PR), k) Lsfc for A1-A5 (PR), l) Lsfc for C1-C5 (PR).

Length-scale analysis, shown in Fig. 7e-l, presents the measured surface profile at different length sections (Fig. 8), and area-scale analysis, plotted in Fig. 7a-d, relies on covering the surface with triangular tiles of different sizes (Fig. 9).

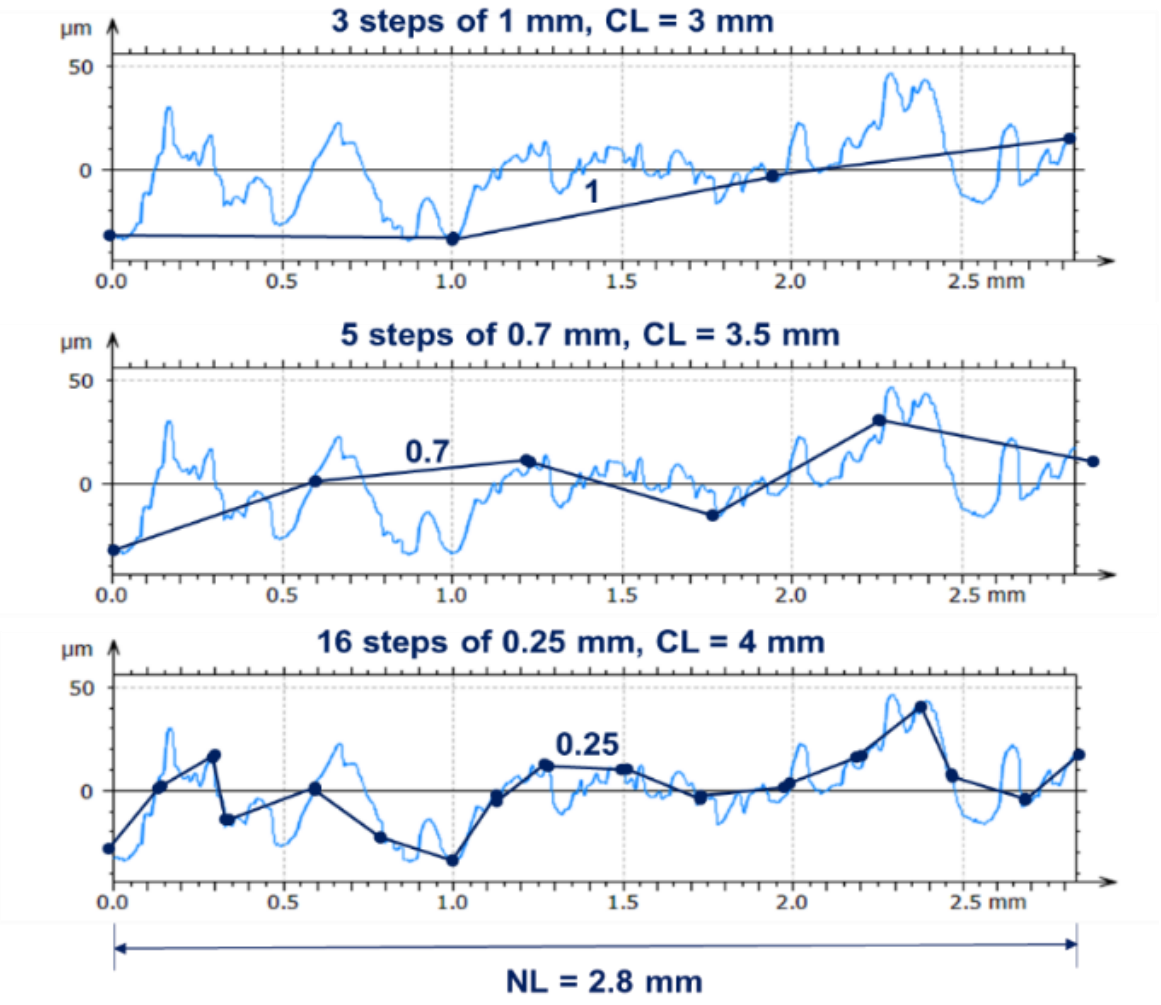

Fig. 8. Example visualization of covering the profile with length sections of different dimensions (length-scale analysis). Note: CL – calculated length, NL – nominal length.

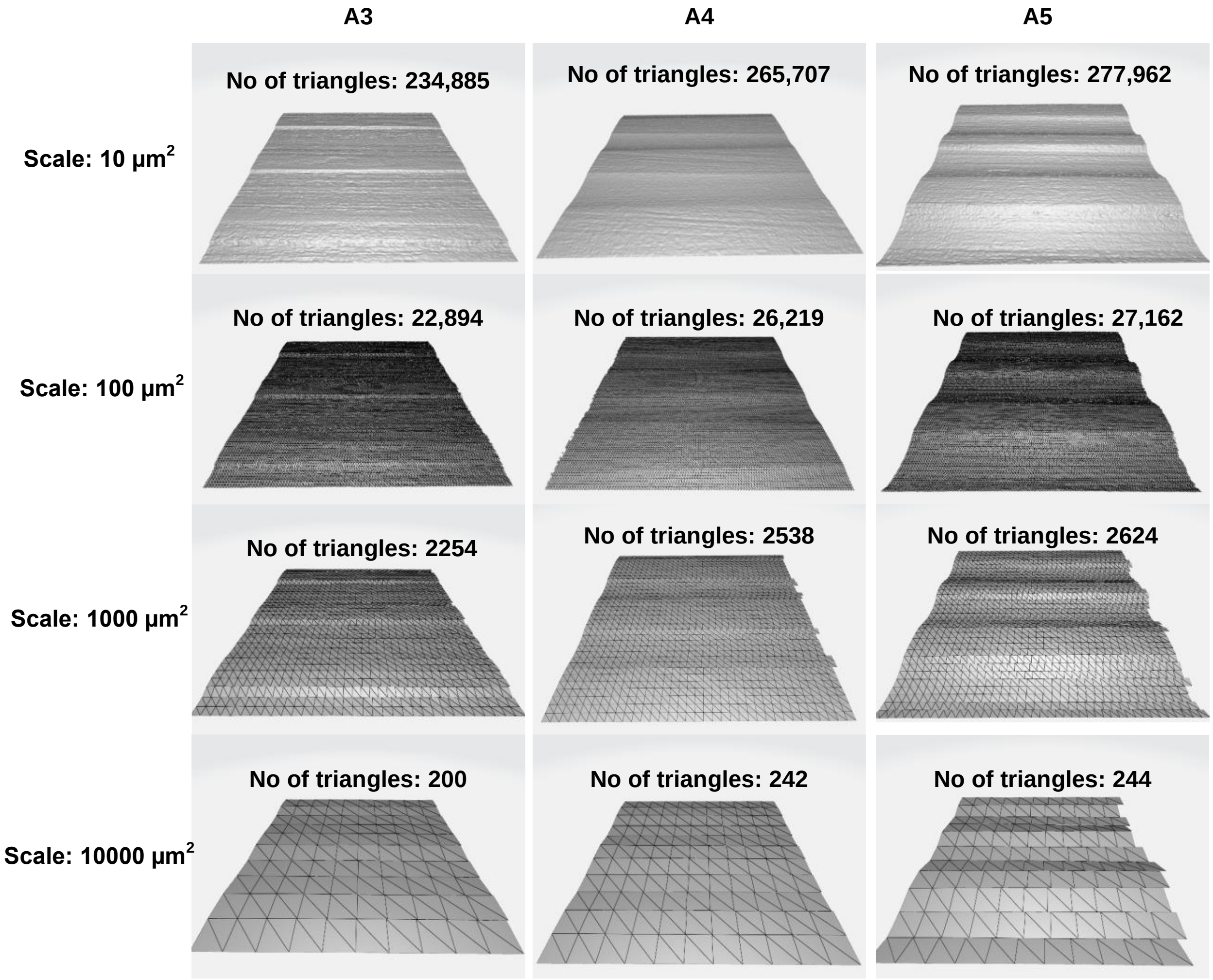

Fig. 9. Example visualization of covering textured surfaces with triangular tiles at various scales (area-scale surface complexity analysis).

Dynamic contact angle hysteresis with advancing and receding contact angles are measured for aluminum alloy (A1-A5) in Fig. 10, and ceramics (C1-C5) in Fig. 11. The experiments consider the forced movement of the liquid droplet in the directions perpendicular (PR) and parallel (PL) to the surface grooves, indicating entrapment of lubricant in the groove valleys and surface drainage. Here, the sliding angle determines the surface tilt angle, causing the liquid droplet to move by changing the contact area between the droplet and the surface. Contact angle recording stopped when the contact area started to move and the droplet crossed the topographic peak. The last point of the curve on the dynamic contact angle hysteresis versus time graph refers to the sliding angle. The shape of the droplet represents the shift in the liquid-surface contact area, along with the corresponding advancing and receding contact angles, is shown in Fig. 12.

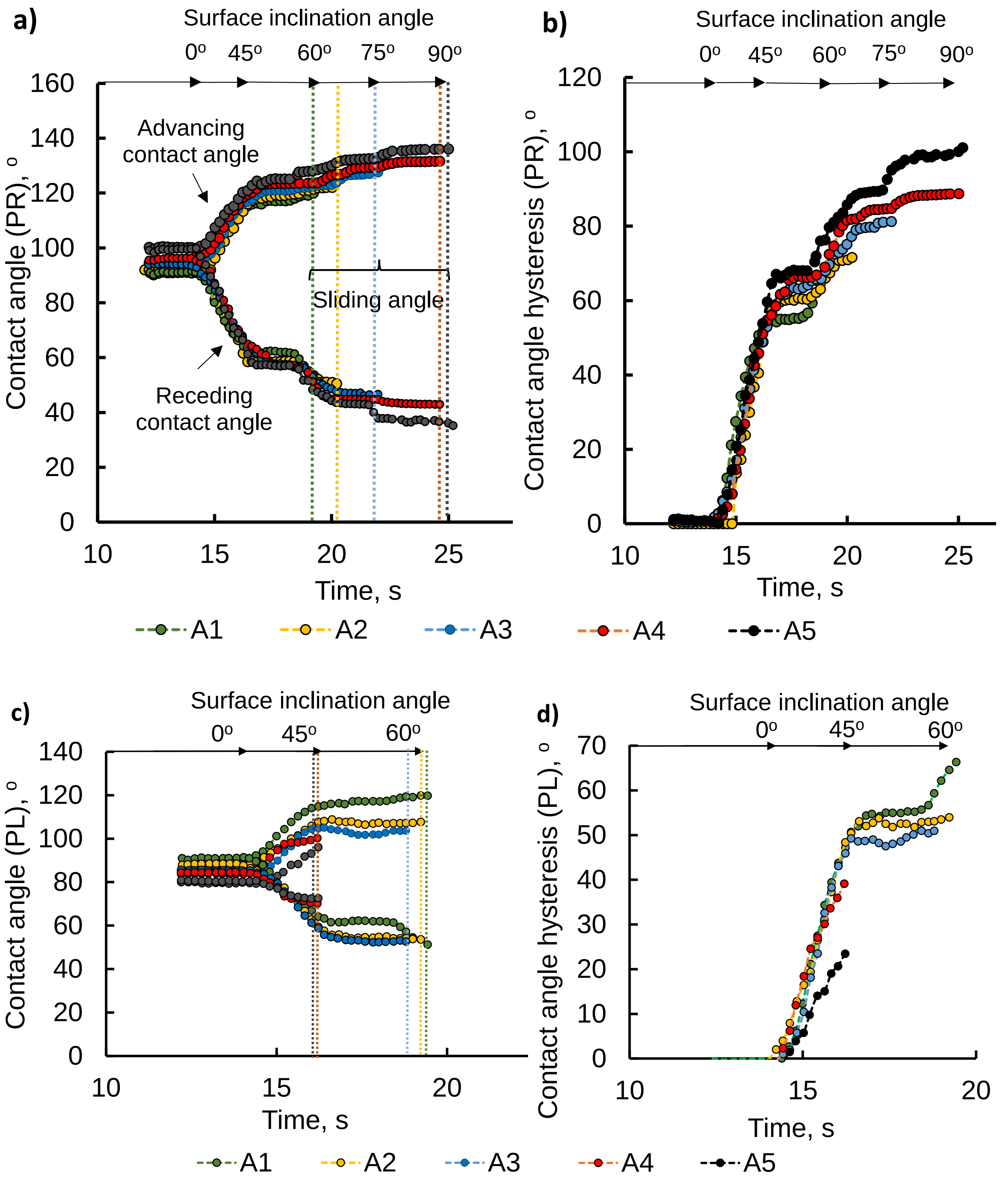

Fig. 10. Dynamic wettability of aluminum alloy surfaces (A1-A5): a) advancing, receding, sliding angles - wettability perpendicular to the grooves (PR), b) dynamic hysteresis of the contact angle perpendicular to the grooves (PR), c) advancing, receding, sliding angles - wettability parallel to the grooves (PL), d) dynamic hysteresis of the contact angle parallel to the grooves (PL).

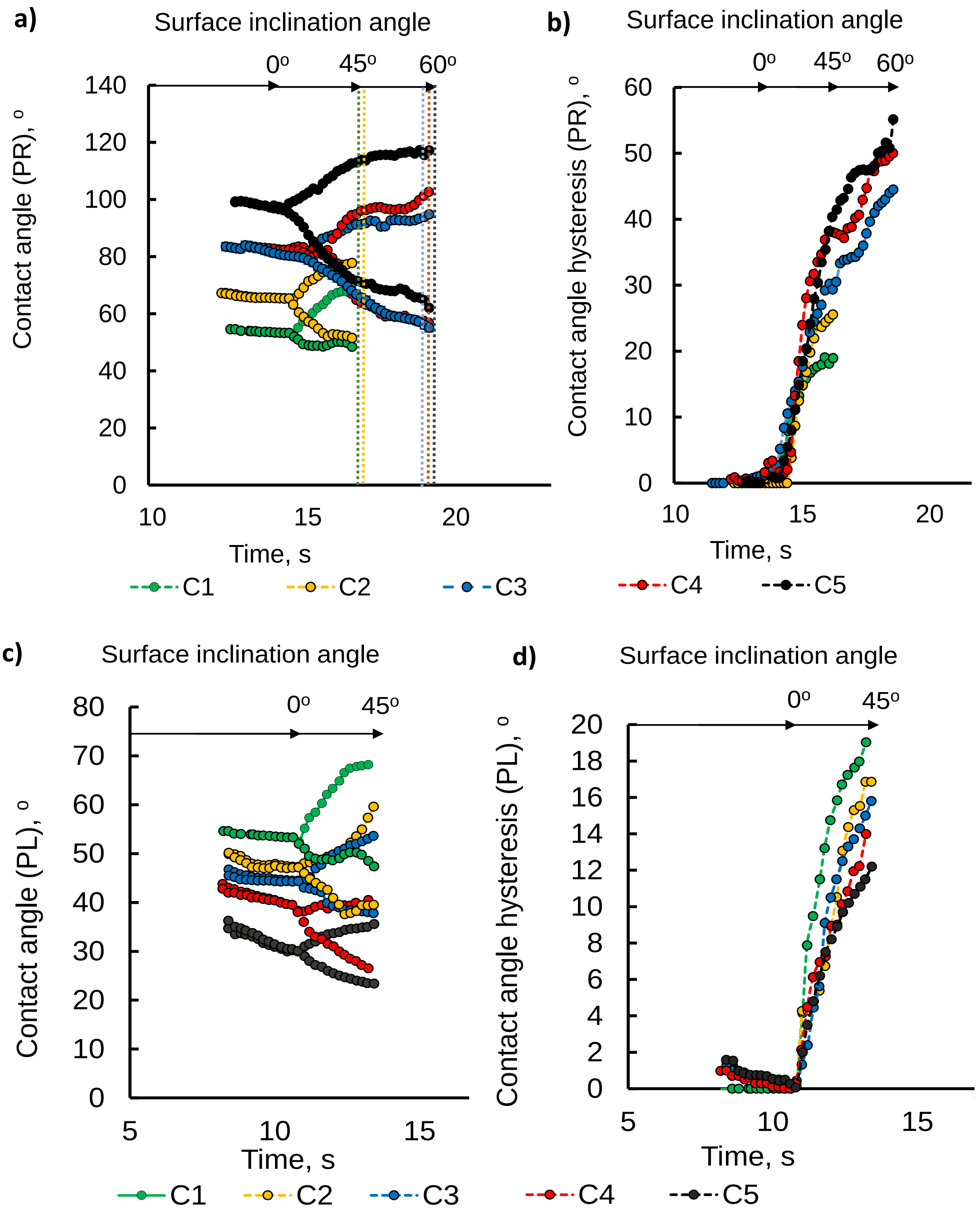

Fig. 11. Dynamic wettability of ceramics (C1-C5): a) advancing, receding, sliding angles - wettability perpendicular to the grooves (PR), b) dynamic hysteresis of the contact angle perpendicular to the grooves (PR), c) advancing, receding, sliding angles - wettability parallel to the grooves (PL), d) dynamic hysteresis of the contact angle parallel to the grooves (PL).

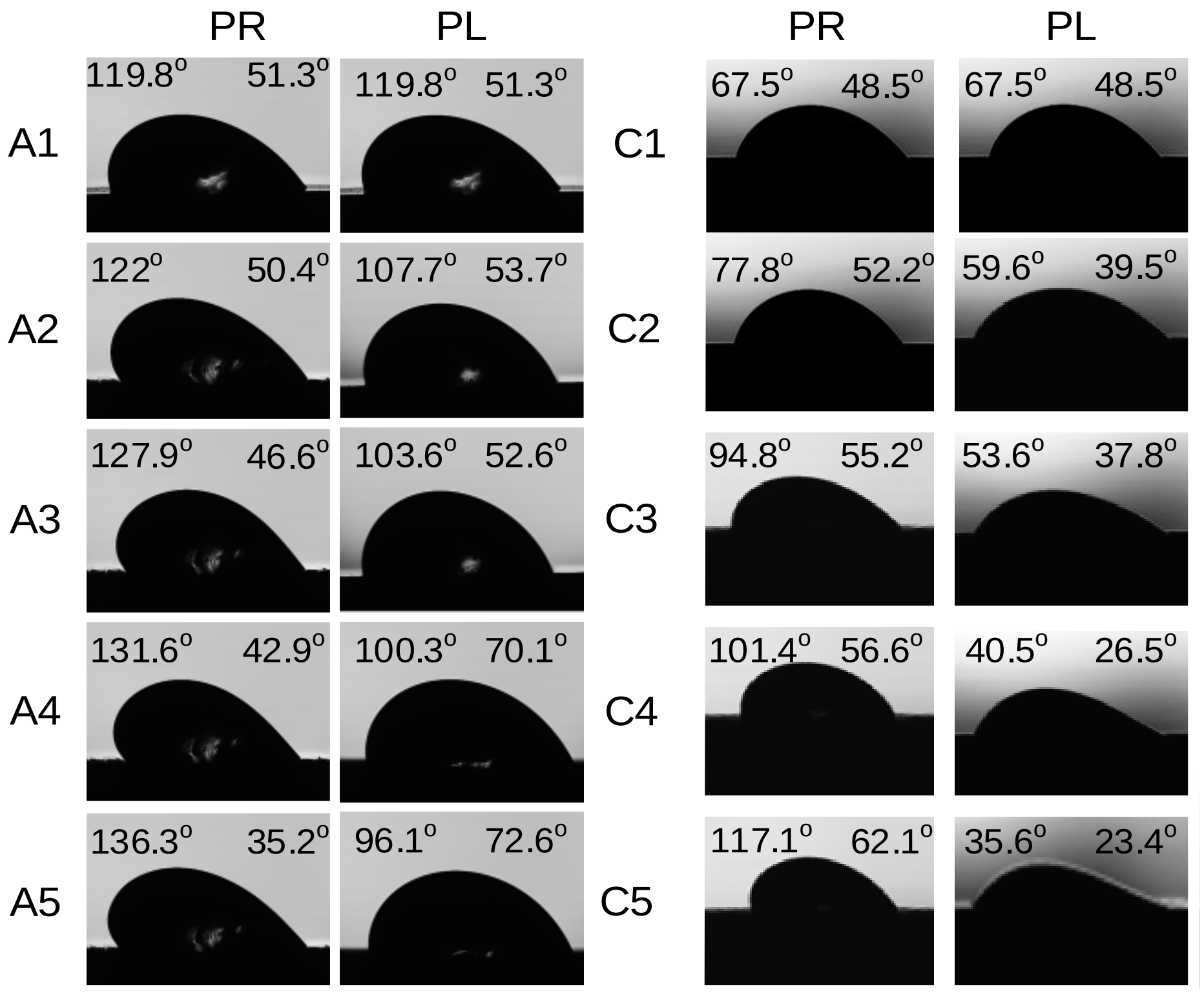

Fig. 12. Representative liquid drops of surfaces A1-A5, C1-C5 at the stage of crossing the topographic peaks in the direction perpendicular to the grooves (PR) and at the moment of sliding along the grooves (PL).

The Pearson correlation coefficients (r) between dynamic contact angle hysteresis CAH and topographic complexity at various observation scales are calculated and presented in Fig. 13. Three different complexity representations were implemented: area-scale complexity Asfc (Fig. 13a), length-scale complexity Lsfc from the length of profiles parallel to the grooves (Fig. 13b), and length-scale complexity Lsfc from the length of profiles perpendicular to the grooves (Fig. 13c). Profiles perpendicular to the grooves better depict the finest and coarsest geometries of topographic ridges and valleys. Correlation coefficients r between CAH and Asfc or Lsfc are presented relative to measurement observation scales, i.e., from 0.58 $\mu m^2$ to 1,318,420 $\mu m^2$ in area-scale and from 1.32 μm to 1,624 μm in length-scale analyses. Due to the wide range of scale values, the x-axis is presented on a logarithmic scale. Here, the finest scales represent the most detailed features of surface microgeometry, while the coarsest scales represent the largest aspects of texture, such as waviness. Correlation plots show the relationships between CAH and Asfc or Lsfc at different surface inclination angles. Furthermore, correlations were calculated for surfaces on which the drops maintained a line of contact with the surface up to a given inclination angle. Correlation coefficients close to +1 indicate the best directly proportional relationship between topographic features of a given scale and the dynamic contact angle hysteresis, while coefficients close to -1 also indicate the best relationship, but inversely proportional. Whereas r around 0 means weak or no correlation of these variables.

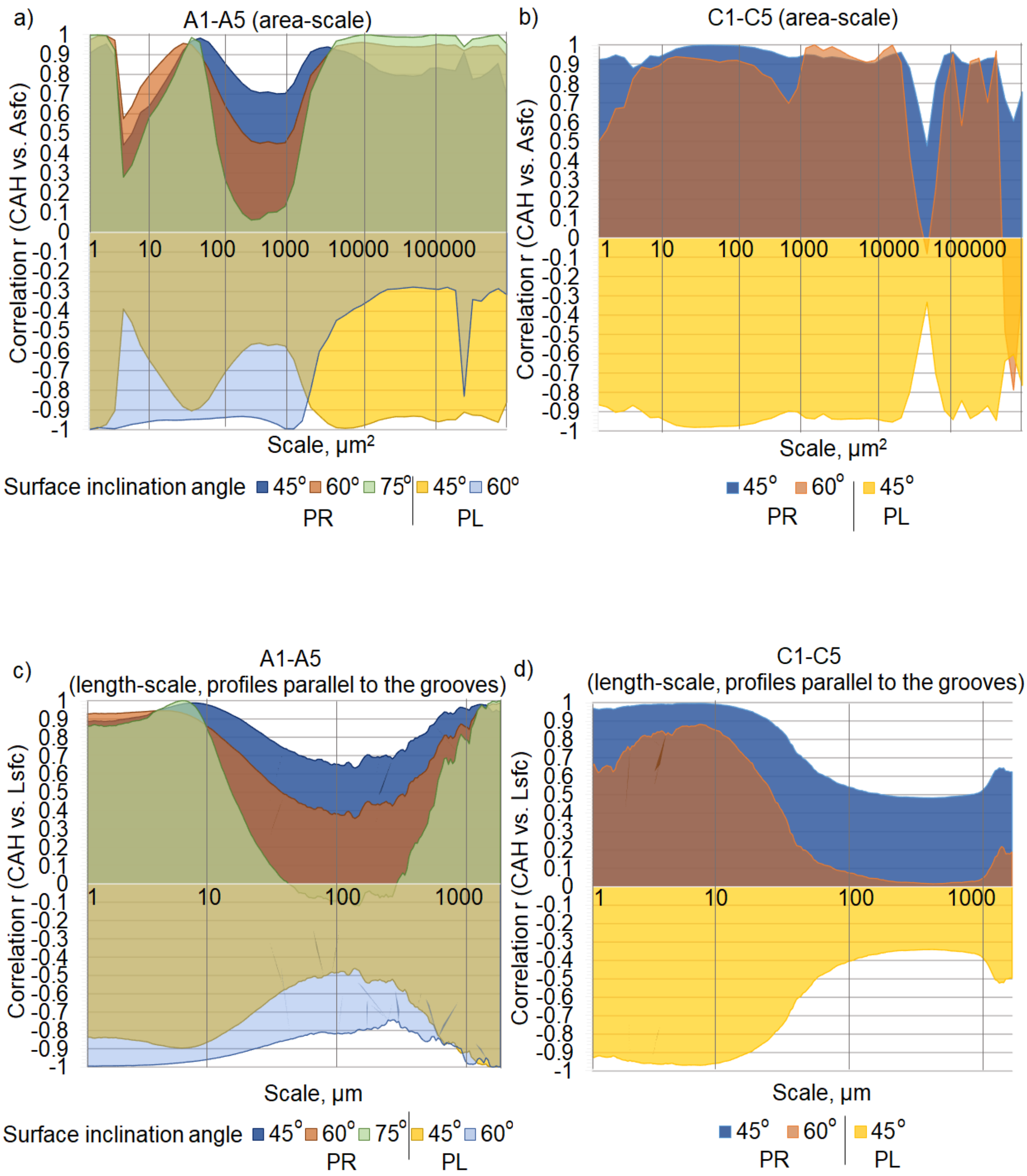
a)
A1-A5 (area-scale)
Correlation r (CAH vs. Asfc)
Scale, µm²
Surface inclination angle 45° 60° 75° 45° 60°
PR
PL
b)
C1-C5 (area-scale)
Correlation r (CAH vs. Asfc)
Scale, µm²
45° 60° 45°
PR
PL
c)
A1-A5
(length-scale, profiles parallel to the grooves)
Correlation r (CAH vs. Lsfc)
Scale, µm
Surface inclination angle 45° 60° 75° 45° 60°
PR
PL
d)
C1-C5
(length-scale, profiles parallel to the grooves)
Correlation r (CAH vs. Lsfc)
Scale, µm
45° 60° 45°
PR
PL

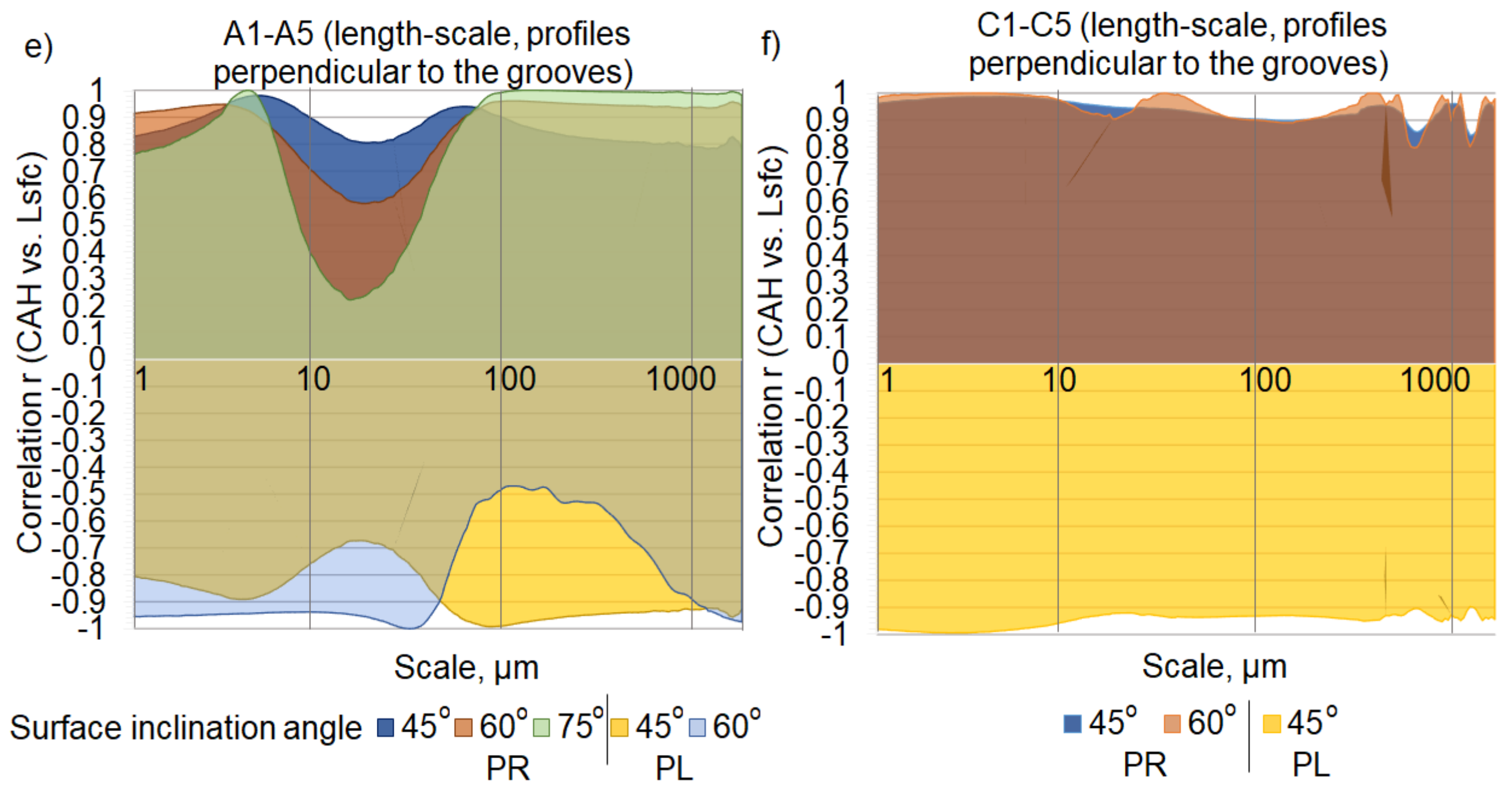

Fig. 13. Multiscale correlations of topographic complexity (area-scale topographic complexity Asfc, length-scale topographic complexity Lsfc) and dynamic contact angle hysteresis (CAH) for textured aluminum alloy (A1-A5) and ceramic (C1-C5) surfaces inclined in the direction perpendicular (PR) and parallel (PL) to the grooves: a) area-scale correlations of surface A1-A5, b) area-scale correlations of surface C1-C5, c) length-scale (profiles parallel to the grooves) correlations of surface A1-A5, d) length-scale (profiles parallel to the grooves) correlations of surface C1-C5, e) length-scale (profiles perpendicular to the grooves) correlations of surface A1-A5, f) length-scale (profiles perpendicular to the grooves) correlations of surface C1-C5.

## 4. Discussion

Area- and length-scale complexities in multiscale geometric topographic characterizations (Fig. 7) complement conventional ISO 25178 parameters. While conventional ISO parameters describe surfaces with a single value representative of the entire range of measurement scales, the multiscale parameters (area- and length-scale complexities) characterize surfaces at different observation scales, separating the finest from the coarsest surface geometries. Although hybrid ISO parameters, Sdr and Sdq, are sensitive to the finest scales of surface geometry, the height-based ISO parameters, Sa, Sq, Ssk, Sku, Sp, Sv, Sz, are based on a large range of scales. Despite these generalities, ISO parameters calculated directly from topographic heights are not designed to describe topography at various, specific observation scales. This function is assumed by multiscale parameters, such as those based on covering the surface with self-similar triangular tiles or profiles of various scales (sizes). These parameters are: relative area, relative length, area-scale complexity and length-scale complexity. This enables characterization of surfaces from the finest to the coarsest scales (in the measurement range from sampling distance to field of view). Multiscale geometric parameters are more comprehensively linked to surface functional characteristics than conventional ISO parameters. This is due to the presence of size-specific microgeometries that correlate better with the functional features of the surface than others. Therefore, specific scales can be identified that best correspond to the surface functionalities.

It is important to situate the study within established theoretical frameworks [66]. Richardson first raised the classic problem of measuring Britain's coastline, which later informed Mandelbrot's work on fractals [67]. Mandelbrot initiated the studies on fractal geometry [68]. Their studies demonstrated that the logarithm of the measured length of irregular lines can change linearly with the logarithm of the observation scale. This relationship allows the coastline of Britain to be distinguished from others, such as Europe, through the slope of the resulting plot. These findings laid the groundwork for the concept of fractal dimensions, providing a means to quantify the geometric complexity of irregular, highly convoluted lines.

The findings presented in these studies extend existing theories of fractal-like surfaces by showing that the aspects of surface complexity relevant to dynamic wetting are not adequately captured by the single-scale or scale-invariant fractal descriptors traditionally used in surface metrology [69]. Classical fractal models generally assume that surface roughness scales uniformly, implying that a single fractal dimension (Das) is sufficient to characterize topographic complexity [70]. In contrast, scale-sensitive fractal analysis (aka multiscale geometric analysis) demonstrates that the relationship between topographic complexity and dynamic wetting is strongly scale dependent. This indicates that wetting is governed not by global fractal roughness, but by multiscale geometric surface features that interact with the liquid interface in distinct ways across different observational scales.

The particular applicability of multiscale geometric analyses is dedicated to the wettability and lubrication of anisotropic surfaces. Wettability strongly correlates with surface texture. Grooved surfaces are characterized by both roughness and waviness. Multiscale geometric analyses indicate the scales of best correlation between wettability and the size of topographic features. Here, the relationships of dynamic surface wetting with the finest microgeometries - roughness, and the coarsest - waviness were identified. Moreover, multiscale geometric analyses identify topographic scales that are dependent on the displacement of the liquid-surface contact line.

In principle, the likelihood of discovering functional correlations depends on characterizing the topography at appropriate scales with geometric pertinence. Both the scale and the nature of the topographic interactions should be considered [1]. A reductionist approach suggests that macroscopic phenomena, such as wetting, can be considered as an agglomeration of discrete interactions at fine scales. Wetting droplets, which might be considered discrete fundamental interactions, also appear to have discrete finite interactions along the wetting line with the solid. The contact line of liquid droplets on irregular surfaces is also irregular. Contact angles along the line vary with the local inclinations on the surface. At sufficiently fine scales, these variations might also not be observable and appear smooth. Sufficiently large variations overcome the surface tension's smoothing of droplets' surfaces, which pulls droplets into smooth shapes above the surface. This agglomerates discrete fundamental interactions at the interface into a measurable, larger-scale phenomenon.

The surface topographic area complexity (Asfc) of both hydrophobic, aluminum alloy (Fig. 13a), and hydrophilic, ceramic (Fig. 13b), surfaces showed the best linear correlations ($r > 0.9$) with the dynamic contact angle hysteresis (CAH) at the fine scale of 28 μm$^2$ and ($r > 0.8$) the coarser waviness scale of 296,335 μm$^2$ when forcing the droplet motion in the direction perpendicular to the grooves. Similar conclusions are observed in the direction parallel to the grooves, but with a slightly lower correlation strength ($r < -0.8$), the negative value of which indicates an inversely proportional relationship between CAH and Asfc. Hydrophobic surfaces are characterized by $r > 0.9$ or $r < -0.9$ also on the finest microroughness scales close to the measurement sampling distance, in the range of 0.58-1.43 μm$^2$. On hydrophobic surfaces, the liquid is difficult to slide, and it is easier to find relationships between fine microgeometry and the behavior of a liquid droplet.

The length-scale complexity (Lsfc) of hydrophobic surfaces, considering profiles in directions perpendicular (Fig. 13e) and parallel (Fig. 13c) to the grooves, correlates similarly with the dynamic hysteresis of the contact angle of a drop moving in directions perpendicular and parallel to the grooves. The linear correlation is strongest for $r > 0.85$ and $r < -0.85$ at the finest and coarsest linear scales, and is 6.9 μm and above 1100 μm, respectively.

The topographic complexity (Lsfc) based on the length of profiles perpendicular to the grooves of hydrophilic surfaces correlates strongly ($R^2 > 0.9$) with the dynamic contact angle hysteresis of a droplet moving perpendicular to the grooves over almost the entire scale range. Both the finest microgeometries and surface waviness interact with dynamic wetting to a similar degree (Fig. 13f). However, (Lsfc), based on the length of profiles parallel to the grooves, correlates strongly with dynamic contact angle hysteresis only at fine scales around 6.9 μm (Fig. 13d). The waviness measured in length-scale profiles correlates with the dynamic contact angle hysteresis to a level of approximately $R^2 = 0.5$

The dynamic contact angle hysteresis is an important factor determining the wettability and lubrication of the functional surfaces. The dynamic contact angle hysteresis mainly depends on the surface roughness and waviness, but also the hydrophobicity or hydrophilicity of the material is considered important [71]. In many practical applications, surfaces are subjected to spontaneous or forced wetting [72]. In this work, the forced change of surface inclination angle with the deposited droplet affected the lubrication of the surface (Fig. 3). Anisotropic, grooved textures are characterized by anisotropic wetting, easier in direction along the surface valleys (Fig. 10c, Fig. 11c), but also in the more difficult direction through the surface ridges (Fig. 10a, Fig. 11a). Surface texturing can affect the entrapment of lubricant in surface valleys and therefore the ability to lubricate the contact surfaces over time. Surface texture also determines the drainage of liquid away from the surface contact zone.

Physical factors influence dynamic wettability on grooved surfaces. Groove geometry and anisotropy determine contact line pinning and de-pinning, directly affecting advancing and receding contact angles, and consequently contact angle hysteresis [73]. Groove orientation relative to droplet motion controls directional surface wetting [74]. The combined effects of surface tension and gravity during droplet motion on inclined textured surfaces determine contact angle hysteresis [75]. Capillary forces on grooved surfaces can arise from groove geometry, which controls contact line pinning and the directional dynamics of surface wetting [76]. Quantitative characterization of these physical parameters is essential for understanding droplet dynamics on textured surfaces.

In these studies, abrasively textured surfaces showed greater area- and length-scale complexity (Asfc, Lsfc) at certain scales (Fig. 6), corresponding to higher height (Sa, Sq, Sp, Sv, Sz) and hybrid (Sdq, Sdr) ISO parameters. The dynamic contact angle hysteresis (CAH) is closely dependent on the orientation of the topographic grooves. Larger area- and length-scale complexities determine a larger dynamic contact angle hysteresis of the drop moving in the direction perpendicular to the grooves (Fig. 10b, Fig. 11b) and, conversely, a smaller CAH in the direction parallel to the grooves (Fig. 10d, Fig. 11d). Larger advancing contact angles are required to overcome larger topographic ridges, resulting in greater CAH. Larger topographic valleys, on the other hand, facilitate droplet flow along them, indicating a capillarity effect and resulting in smaller droplet shape changes (differences between advancing and receding contact angles). These assumptions were confirmed for both hydrophilic and hydrophobic materials. Although the hydrophobic surfaces, compared to hydrophilic surfaces, were characterized by approximately twice the CAH of the droplet moving in both directions perpendicular and parallel to the grooves.

Butt et al. [22], in a review paper, consider surface roughness and heterogeneity as the main factors influencing dynamic contact angle hysteresis. Surface features such as grooves cause pinning of the contact line, leading to a larger advancing angle (Fig. 12). Smaller advancing angles are observed for droplet displacements along the grooves, which is also due to easier movement in this direction and the lack of transition barriers. If the drops are forced to move in a direction perpendicular to the grooves, the advancing angle increases with each subsequent transition barrier. The higher the barrier, the greater the advancing contact angle. The trend of changing the advancing contact angle in the direction perpendicular to the grooves is different for hydrophobic and hydrophilic materials. For hydrophobic materials, the more complex the surface texture and the larger the grooves, the greater the advancing angle, while for hydrophilic materials, an inverse trend can be observed. Textured aluminum alloy, which is hydrophobic, show stronger pinning of liquid droplets than hydrophilic textured ceramic surfaces. This difference stems from the higher surface free energy of ceramics, which promotes liquid spreading [77]. Consequently, although texturing ceramics increase droplet pinning, the effect is less pronounced than in materials with lower surface free energy, such as aluminum alloys. Groove texturing on aluminum alloy surfaces, with Sa values ranging from 0.43 to 18.83 μm and Sz values from 12.27 to 73.87 μm, resulted in an increase in the advancing contact angle from 119.8° to 136.3°, respectively.

The parameter that determines the surface inclination angle in order to induce lubricant movement is the sliding angle. In the case of both textured hydrophobic and hydrophilic surfaces, this trend is heavily dependent on the sliding direction. The opposite trend was observed for PR (Fig. 10a, Fig. 11a) and PL (Fig. 10c, Fig. 11c) directions on surfaces. Although hydrophilic surfaces are characterized by faster spreading of lubricant on the surface, and therefore a smaller sliding angle.

While surface texture is an essential determinant of wetting and lubrication, the type of material is also important. Legrand et al. [78] analyzed the anisotropic wettability of hydrophobic polymeric materials with the same topographic characteristics, for which different dynamic contact angle hysteresis values of 400% were obtained. Ijaola et al. observed that low surface energy materials lead to surface hydrophobicity [79]. In this work, comparison of the dynamic contact angle hysteresis for a hydrophobic material (aluminum alloy) and a hydrophilic material (ceramics) demonstrated the importance of chemical dependencies in the modeling of the contact angle. Similar values of conventional ISO and multiscale geometric parameters between both surfaces, ceramics and aluminum alloy, can lead to different dynamic contact angle hysteresis.

The literature states that the difficulties in wetting studies are the multitude of factors determining wetting [80]. The static contact angle is described primarily by the Young, Wenzel, and Cassie-Baxter models. Analysis of static contact angles with these models is widely used and considered reliable [81]. The Wenzel and Cassie-Baxter models allow for the consideration of surface roughness, but do not include any other factors. Chau et al. [80] noted that the key in analyzing surface wettability is not the strict introduction of measurement data into ideal mathematical models of wetting, but rather the characterization of the surface in terms of topographic and chemical characteristics in static and dynamic interactions with the behavior of liquids [72]. The dynamic wetting is especially important from the application point of view [82]. In particular, the novelty of these studies incorporating multiscale correlations of topographic complexity and dynamic contact angle hysteresis provides insight into the scales of microgeometries specifically interacting with dynamic wettability. This is of general importance in most engineering systems operating in a solid-liquid system involving the motion of a liquid, for example, tribological systems.

## 5. Conclusions

These studies have demonstrated a multiscale measurement-based procedure for finding relationships between the topographic complexity of hydrophilic and hydrophobic anisotropic surfaces and the dynamic contact angle hysteresis, determinant of dynamic wettability. From the investigations presented in these studies, the following conclusions can be drawn:

- Multiscale geometric parameters are more closely related to surface functional features than conventional ISO 25178 parameters, as they identify specific scales of surface geometry that best represent surface functionalities.
- Multiscale geometric analyses establish correlations from the finest (microroughness) to the coarsest (waviness) surface geometries with dynamic wettability.
- The surface topographic complexity of both hydrophobic and hydrophilic surfaces showed the best linear correlation ($R^2 > 0.9$) with dynamic contact angle hysteresis at a fine roughness area-scale of 28 $\mu m^2$ and slightly worse linear correlation ($R^2 > 0.8$) at a coarse waviness area-scale of 296,335 $\mu m^2$.
- Area-scale complexity correlates better with dynamic contact angle hysteresis in the direction perpendicular to the grooves ($r > 0.9$) than in the direction parallel to the grooves ($r < -0.8$).
- Area-scale complexities of hydrophobic surfaces correlate strongly ($R^2 > 0.9$) also at the finest scales near the measurement sampling distance, in the range of 0.58-1.43 $\mu m^2$; therefore, even the finest microgeometries have an impact on the surface hydrophobicity.
- The length-scale complexity of hydrophobic surfaces correlates strongly with the dynamic contact angle hysteresis ($R^2 > 0.85$) at the finest and coarsest linear scales, and is 6.9 $\mu m$ and above 1100 $\mu m$, respectively.

- The topographic complexity of hydrophilic surfaces correlates strongly ($R^2 > 0.9$) with the dynamic contact angle hysteresis of a droplet moving perpendicular to the grooves over almost the entire scale range; however, a droplet spreading along the grooves correlates strongly only at fine scales around 6.9 μm.
- Greater area- and length-scale complexities determine a larger dynamic contact angle hysteresis of the drop moving in the direction perpendicular to the grooves and, conversely, a smaller dynamic contact angle hysteresis in the direction parallel to the grooves.
- The abrasive, directional surface texturing with a self-sharpening material with a pyramidal structure shapes the surface's topographic characteristics and thus models dynamic wettability.
- Texturing hydrophilic surfaces can change their functional properties to hydrophobic ones.
- Dynamic contact angle hysteresis depends on the roughness and waviness of the surface, but also on the hydrophobicity and hydrophilicity of the material.

Correlations between dynamic contact angle hysteresis and topographic complexity offer applicability in systems involving liquid–solid interactions, including self-cleaning surfaces, heat exchangers, and marine hull coatings. Particularly, in surface engineering, this contributes to the design of tribological systems that operate under hydrodynamic lubrication conditions. This approach identifies the measurement scales that best reflect liquid–solid interactions, leading to more reliable assessments of functional surface properties. It also enables refinement of measurement procedures and the development of more consistent approaches for characterizing functional surfaces.

Acknowledgements

This research was funded by the National Science Centre, Poland, under the program MINIATURA 7, grant number: 2023/07/X/ST8/01233. K.J. Kubiak would like to acknowledge the support provided by the UK EPSRC grant TRENT (EP/S030476/1).

Authorship contribution statement

Katarzyna Peta: Writing – review & editing, Writing – original draft, Visualization, Validation, Supervision, Software, Resources, Project administration, Methodology, Investigation, Funding acquisition, Formal analysis, Data curation, Conceptualization. Krzysztof J. Kubiak: Writing – review & editing, Writing – original draft, Visualization, Validation, Supervision, Software, Resources, Project administration, Methodology, Investigation, Funding acquisition, Formal analysis, Data curation, Conceptualization. Christopher A. Brown: Writing – review & editing, Writing – original draft, Software, Methodology, Formal analysis.